\documentclass{optica-article}
\journal{opticajournal} 
\articletype{Research Article}

\begin{document}

\title{Impact of Nitrogen-Vacancy Color Centers\\ on the Optical Kerr Effect in Diamond:\\ A Femtosecond Z-Scan Study}

\author{Wojciech Talik, Mariusz Mrózek, Adam M. Wojciechowski and Krzysztof Dzierżęga\authormark{*}}

\address{Marian Smoluchowski Institute of Physics, Jagiellonian University, ul. Łojasiewicza 11, 30-348 Kraków,  Poland}

\email{\authormark{*}krzysztof.dzierzega@uj.edu.pl} 


\begin{abstract*} 
This study explores nonlinear optical effects in diamond crystals doped with nitrogen-vacancy (NV) centers, focusing on the optical Kerr effect in the infrared spectral region. By employing the Z-scan technique with 230 fs laser pulses at a wavelength of 1032~nm, we investigate the nonlinear refractive index of high-purity and NV-containing diamond crystals at NV concentrations of 0.3 ppm and 4.5 ppm. The results demonstrate a pronounced anisotropy in the nonlinear refractive index, which exhibits a four-fold rotational symmetry, with its amplitude reduced by the NV concentration. The reduction in non-linear susceptibility is linked to the negative contribution of NV centers, modeled as two-level systems upon a strong, laser field, red-detuned relative to their resonance frequency. 
These findings provide new insights into the interplay between NV centers and nonlinear optical properties of diamond, offering potential pathways for tunable nonlinear optics applications.
\end{abstract*}


\section{Introduction}
Nitrogen-vacancy (NV) centers in the diamond are crystallographic defects that form when a nitrogen atom replaces a carbon atom in the diamond lattice, accompanied by a neighboring vacancy (a missing atom) \cite{doherty_2013}. These defects are responsible for the creation of additional energy levels within the diamond band gap (see Figs.~\ref{Fig: crystals parameters}(b,c)), enabling optical excitation and fluorescence in the visible and near-infrared, as well as spin manipulation using microwave fields. Owing to their long spin coherence times and stability even at room temperature \cite{shinei_2022}, NV centers have been widely applied in high-precision magnetometry \cite{acosta_2009}, temperature sensing \cite{choe_2018}, and pressure detection \cite{lesik_2019}. Their sensitivity to minute changes in environmental conditions has made them essential for applications such as nanoscale biological imaging \cite{hall_2013}, quantum sensing \cite{rizzato_2023}, and quantum information processing \cite{xu_2019}. Diamond atomic force microscopy (AFM) tips embedded with NV centers or their ensembles have enabled magnetic imaging with nanometer-scale spatial resolution \cite{ardagozen_2014}. Furthermore, in biological systems, diamonds containing NV serve as highly effective fluorescence markers that are not toxic and do not photobleach, allowing non-invasive diagnostic applications \cite{balasubramanian_2014, kolodziej_2024}. Recently, NV diamonds have been integrated into optical fibers \cite{filipkowski_2022}, creating localized or distributed sensors capable of operating in extreme environments where conventional technologies usually fail.

Diamond crystals belong to the crystallographic m3m point group and exhibit inversion symmetry. However, the introduction of NV centers locally breaks this symmetry, resulting in a non-zero second-order nonlinear susceptibility, $\chi^{(2)}\neq0$. This symmetry breaking, along with additional energy levels within the diamond band gap, may lead to second-order nonlinear optical (NLO) effects such as second harmonic generation (SHG) and the electro-optic (EO) effect. Moreover, it can also enhance third-order NLO effects, such as the optical Kerr effect (OKE) and two-photon absorption (2PA) in the near-infrared range, the effects studied in this work.

Although significant attention has been devoted to the magnetic properties of diamonds with NVs and their well-established roles in sensing and metrology, their NLO properties have been less explored. 
Relevant early studies include four-wave mixing experiments by Rand \cite{rand_1988} and studies of Sheik-Bahae et al. \cite{sheikbahae_1995}. 
Recently, Motojima et al. \cite{motojima_2019} used Z-scan and pump-probe techniques and demonstrated enhanced OKE and 2PA due to NV centers located near the near-surface layer of the diamond, when excited by the 800~nm 40~fs laser pulses. SHG and SHG-induced cascaded third-harmonic generation in bulk diamond doped with NV, excited by ultrashort femtosecond laser pulses in the spectral range of 1200--1600~nm, was observed by Abulikemu et al. \cite{abulikemu_2021, abulikemu_2022} .

The recent commercial availability of high-quality diamond plates, uniformly doped with NV centers throughout their volume \cite{Element6}, has significantly accelerated research in NV sensing. These diamonds not only provide long spin coherence times (important for applications in sensing and information technology), but also ensure a uniform distribution of NV centers, which was previously unattainable in laboratory-prepared samples. Consequently, researchers now have access to more reliable materials, facilitating the exploration of both established and novel applications ranging from quantum sensing to nonlinear optics.\\

In this study, we explore the impact of NV centers on nonlinear optical effects in diamonds, specifically focusing on the optical Kerr effect in the infrared region. We employ 1032~nm, 230~fs laser pulses at a 1~kHz repetition rate and utilize the Z-scan method in closed-aperture (CA) mode \cite{sheikbahae_1990a}. 
Our findings reveal strong nonlinearity of the refractive index that scales linearly with the laser intensity. We also observe a strong anisotropy in the nonlinear refractive index of the pristine diamond, characterized by a four-fold rotational symmetry when illuminated normal to the (001) surface.
This symmetry persists also in diamonds containing NV, although the magnitude of the index nonlinearity is reduced when the NV density increases. 
The reduction can be attributed to the negative contribution of the third-order susceptibility from NV centers, when treated as two-level systems exposed to a strong laser field red-detuned relative to their generalized resonant transition frequency.

\section{Experiment}
In this study, we examined three commercially available diamond crystals with different properties: a high-purity electronic-grade single crystal and two crystals doped with NV color centers at medium concentration of 0.3 ppm and at high concentration of 4.5 ppm, hereafter referred to as EGSC, MCNV, and HCNV, respectively. The EGSC crystal was supplied by Element Six, while the two others were obtained from Thorlabs. A summary of their key characteristics is presented in Table \ref{tab:sample-characteristics}. The transmission spectra of these crystals were measured in the range 400--1100~nm using a dual-beam UV/VIS/NIR spectrophotometer (SPECORD210 Plus, Analytik Jena) with a spectral bandwidth of 1.0 nm. 
\begin{table}[htb]
\centering
\small
\caption{\bf Characteristics of the investigated diamond crystals.}
\begin{tabular}{ccccccc}
\hline
Designation & N [ppm] & NV [ppm] & Size [$\rm mm^3$]  & Face & Edges  & Supplier \\
\hline
EGSC & $< 0.005$ & $-$ & $\rm 2 \times 2 \times 0.5$ & $(001)$ & $[110]$ & Element Six \\
MCNV & $0.8$ & $0.3$ & $\rm 3\times3\times0.5$ & $(001)$ & $[100]$ & Thorlabs\\
HCNV & $13$ & $4.5$ & $\rm 3\times3\times0.5$ & $(001)$ & $[100]$ & Thorlabs\\
\hline
\end{tabular}
  \label{tab:sample-characteristics}
\end{table}

\begin{figure}[t!]
    \centering\includegraphics[width=15.0cm]{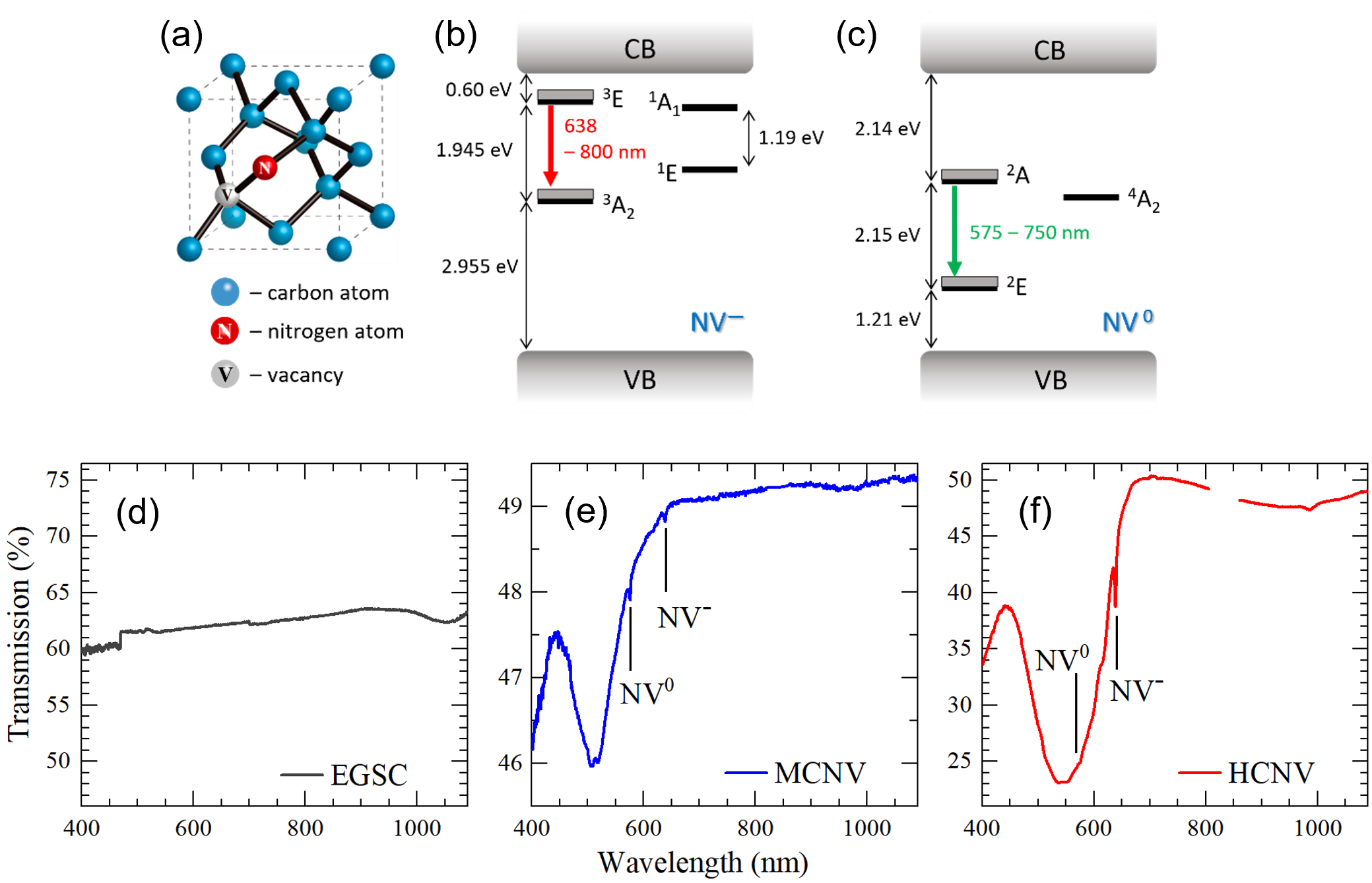}
    \caption{Properties of a diamond crystal containing NV color centers. (a) Model of the crystallographic structure of the NV color center. (b-c) Energy level structure of the negative and neutral nitrogen-vacancy color centers in the band gap of the diamond. (d-f) Transmittance spectra of the three crystal samples studied.}
    \label{Fig: crystals parameters}
\end{figure}

The acquired transmission spectra for all investigated samples are shown in Figs.~\ref{Fig: crystals parameters}(d-f). While the spectrum of an undoped sample EGSC is almost flat, the transmission spectra of MCNV and HCNV display a broad absorption band approximately 2$\%$ to 20$\%$ depth (NV-density dependent), centered at the effective peak wavelength around 520~nm (approx. 2.4~eV). We attribute this strong absorption band to transitions within the electron and phonon structures of the NV centers (see Figs.~\ref{Fig: crystals parameters}(b,c)). Within this absorption band, two narrow zero-phonon lines characteristic of the neutral NV$^0$ at 575~nm and negatively charged NV$^-$ at 637~nm color centers are clearly visible. Furthermore, the pronounced increase in absorption in the blue region of the spectrum (below 460~nm) is attributed to transitions from the NV$^-$ ground state $^3A_2$ to the conduction band. 

\begin{figure}[b!]
    \centering\includegraphics[width=8.0cm]{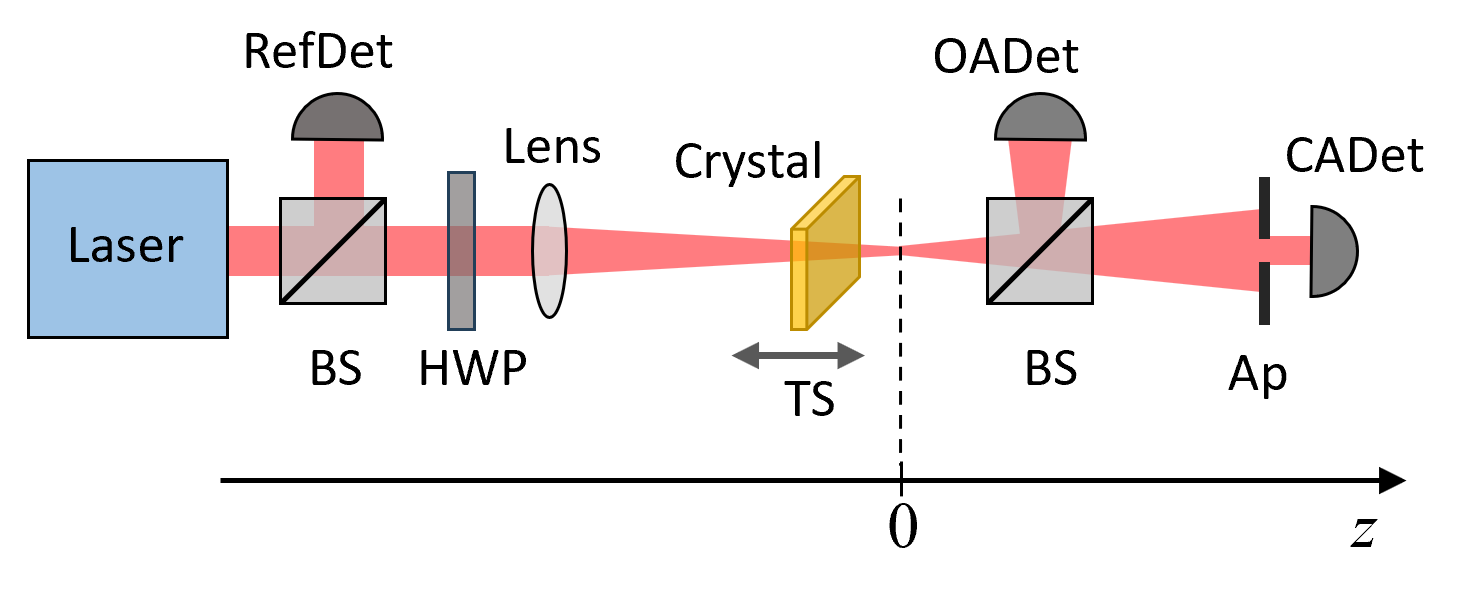}
    \caption{Schematic diagram of the experimental Z-scan setup. RefDet, OADet, and CADet represent the reference, open-aperture, and closed-aperture signal detectors, respectively. BS denotes the beam splitter, HWP the half-wave plate, TS the translation stage, and Ap the aperture.}
    \label{Fig: ExpSetup}
\end{figure}

Nonlinear optical effects, such as the optical Kerr effect and two-photon absorption, were investigated in diamond crystals with NV centers in the infrared transparent region at a wavelength of 1032~nm using the Z-scan technique \cite{sheikbahae_1990a}. Z-scan measurements were performed in closed-aperture (CA) and open-aperture (OA) modes to examine nonlinear refraction and nonlinear absorption, respectively, following a standard experimental setup depicted in Fig.~\ref{Fig: ExpSetup}. A Yb-fiber laser (Jasper Flex, Fluence) was used to generate 230~fs light pulses at 1032~nm (1.201~eV), with a repetition rate of 1~kHz and a maximum pulse energy of 10~$\mu$J. A portion of the laser beam was reflected by a beam splitter to monitor pulse stability, while the transmitted beam was focused by a lens with a focal length of $f=10$~cm into a spot with a diameter of $2w_0=70$~$\mu$m and the Rayleigh range $2z_0=7.3$~mm. 
OA, CA and reference signals were measured using identical large-area photodiodes ( DET100A/M, Thorlabs) connected to a digital oscilloscope (1~GHz, 10~GS/s, Tektronix). The magnitude of the signal, determined as the areas under the waveforms of the transmitted pulses, was stored in the oscilloscope memory, with the OA and CA values normalized to the reference signal. Such normalized signals were averaged over 500 laser shots and subsequently transferred to a computer. This procedure, repeated for each position of the crystal during the Z-scan experiment, significantly improved the signal-to-noise ratio. 

A key limitation of the Z-scan method, in its CA mode, is the movement of the transmitted laser beam across the aperture as the sample is translated along the z-axis, which can degrade the quality of results and significantly reduce the sensitivity. This effect is caused by both sample imperfections and even slight misalignments in the experimental setup. To mitigate this issue, we performed reference measurements at very low pulse energies -- below 30~nJ (7 GW/cm$^2$) in our case -- where no nonlinear effects occure for all $z$-positions. The OA and CA curves measured under these conditions were then used to correct the signals measured at higher laser intensities. Additionally, potential polarization anisotropy was examined by altering the polarization direction of the incident laser beam relative to the crystal axis using a half-wave plate placed in its path.

\section{Results and Discussion}
\begin{figure}[tb!]
    \centering\includegraphics[width=15cm]{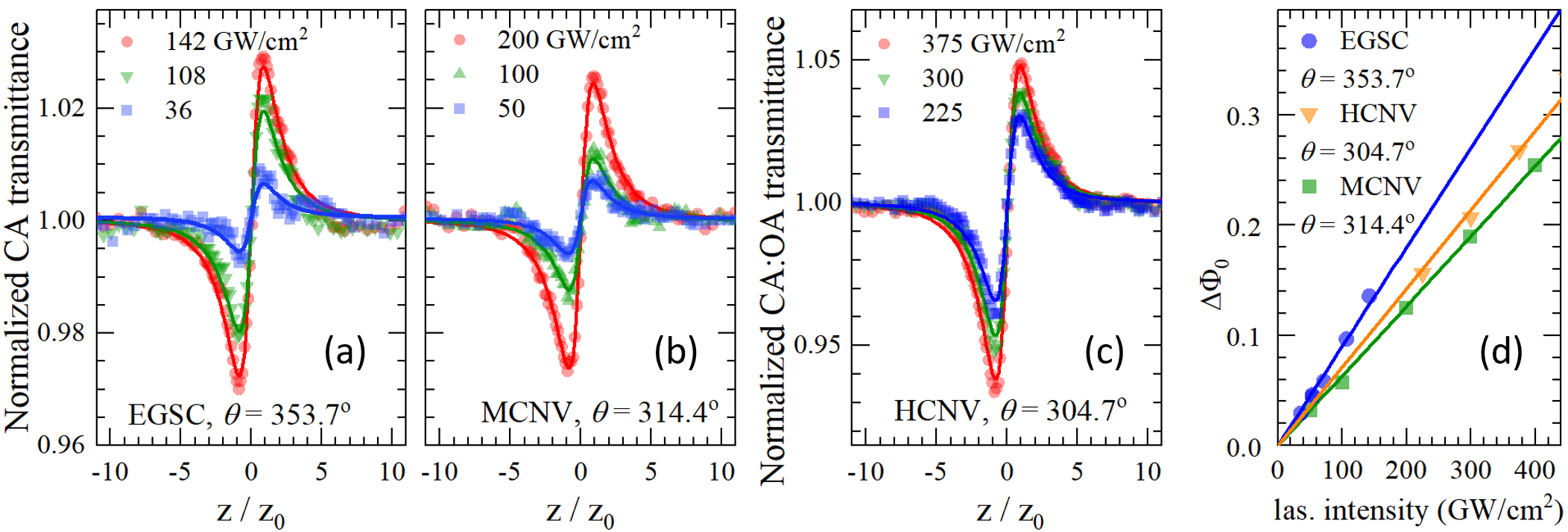}
    \caption{(a-c) Closed aperture (CA) Z-scan results for different laser beam intensities and specific crystal orientations. Solid red lines represent fits to the data using Eq. (\ref{eq:TCA}). (d) The nonlinear phase shift $\Delta \Phi_0$ depending on the laser intensity for different samples with specific crystal orientations.}
    \label{fig:Z-scan results}
\end{figure}

Figure~\ref{fig:Z-scan results} presents the Z-scan results for all studied samples determined at various laser intensities. For the laser intensities studied, both the EGSC and the MCNV exhibit negligible non-linear absorption, so the normalized CA transmittance was directly analyzed. On the contrary, HCNV displayed a saturable two-photon absorption (S2PA) of up to a few percent, prompting a further analysis of the CA:OA transmittance instead. We attribute the observed S2PA effect to the induced two-photon transitions between the ground and excited states of the neutral and negatively charged NV centers. All the acquired CA Z-scan results indicate positive non-linear refraction due to the OKE. This causes the beam diameter to initially increase at the aperture, when the sample is moving towards the photodetector, resulting in a reduced detected signal. In agreement with the basic mechanism of OKE and self-focusing, the overall index of refraction, which depends on the laser intensity $I$, is expressed as $n=n_0+n_2 I$, where $n_0$ is the linear refractive index, and $n_2$ represents the nonlinear refractive index given by
\begin{equation}
    n_2=\cfrac{3}{4}\frac{\mathrm{Re}\left (\chi_{\rm eff}^{(3)} \right )}{n_0^2\; \epsilon_0 c}.
\label{eq:n2 to chi^3}    
\end{equation}
Here, $\chi_{\rm eff}^{(3)}$ is the effective third-order susceptibility which, in general, depends on the symmetry and orientation of the crystal, $\epsilon_0$ is the vacuum permittivity, and $c$ is the speed of light in vacuum.

To quantify the experimental data of our Z-scan measurements, the normalized CA or CA:OA transmittance of the Gaussian laser beam as a function of the sample position $z$ is approximated as in ref. \cite{dzierzega_2021}
\begin{equation}
    T_{CA}(z)=1+\sqrt{\pi}\frac{4(z/z_0) \Delta \Phi_0 - (z^2/z_0^2+3)\Delta\Psi_0}{(1+z^2/z_0^2)(9+z^2/z_0^2)},
\label{eq:TCA}    
\end{equation}
where the nonlinear phase shift is given by $\Delta\Phi_0=kn_2I_0L_{\mathrm{eff}}$, and $k=2\pi/\lambda$ with $\lambda$ representing the laser wavelength while $\Delta\Psi_0$ is responsible for some residual nonlinear absorption. $L_{\mathrm{eff}}=(1-\exp{(-\alpha_0 L)})/\alpha_0$ with $L$ the thickness of the sample and $\alpha_0$ the linear absorption coefficient at the studied wavelength. Finally, $z_0=k w_0^2 /2$
is the Rayleigh range of the laser beam, with $w_0$ being the beam waist and $I_0$ the laser irradiance in the focal plane.  

In Figs.~\ref{fig:Z-scan results}(a-c), the solid lines represent curves fitted to the experimental data using Eq.~(\ref{eq:TCA}).
As shown in Fig.~\ref{fig:Z-scan results}d, at a given polarization of the laser beam, the phase shifts $\Delta \Phi_0$ are directly proportional to the laser intensity within the investigated intensity range, confirming the OKE origin of these observations.\\

\begin{figure}[tb!]
    \centering\includegraphics[width=8cm]{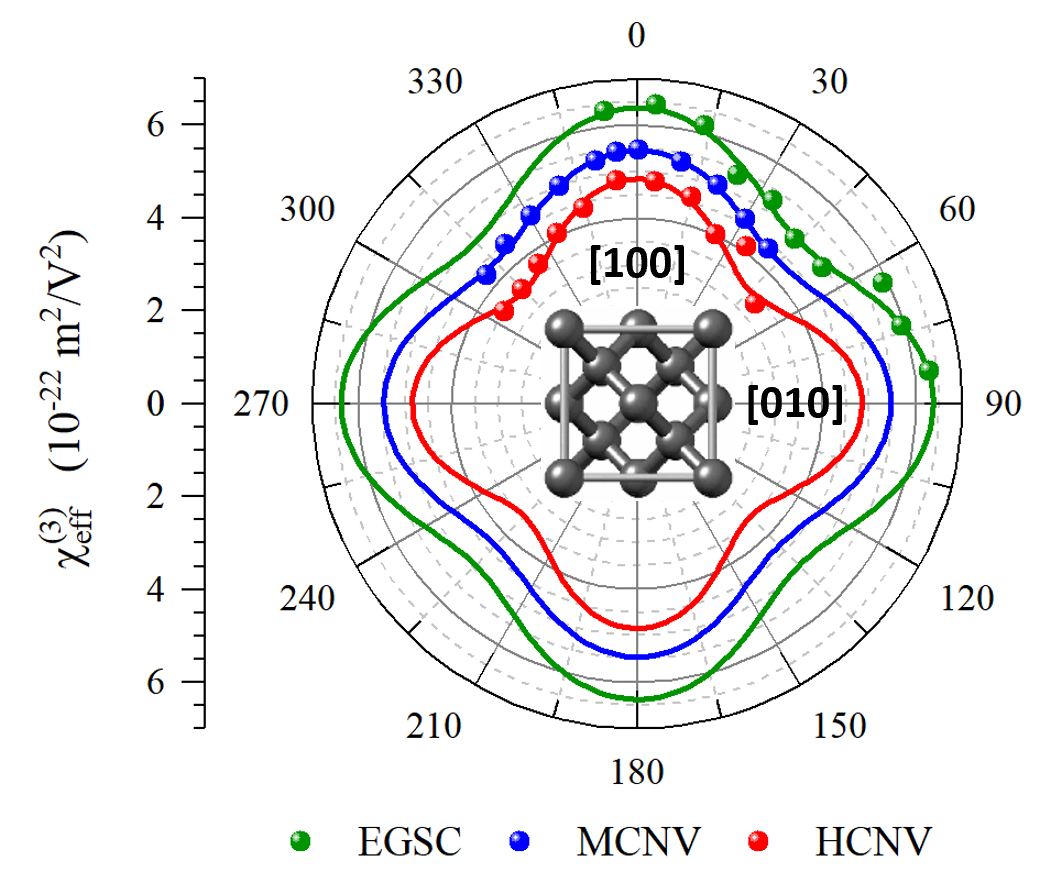}
    \caption{The anisotropy of the third-order nonlinear susceptibility $\chi_{\mathrm{eff}}^{(3)}$ determined for the studied crystals. Angles $\theta = 0^{\mathrm{o}},\;90^{\mathrm{o}}$ correspond to the crystallographic directions [100] and [010], respectively, indicated in the NV model shown in the figure.}
    \label{fig:PolarPlot}
\end{figure}

As mentioned earlier, we studied the nonlinear refractive index as a function of the angle $\theta$ between the electric field vector that characterizes the polarization of the laser light relative to the crystallographic axis [100] of the diamond. The results, which are summarized in Fig.~\ref{fig:PolarPlot},
indicate that all the investigated crystals exhibit anisotropy in their third-order susceptibility $\chi_{\rm eff}^{(3)}$, which is related to nonlinear refraction $n_2$ by Eq.~(\ref{eq:n2 to chi^3}), and which decreases with increasing NV concentration, reaching its lowest value for the HCNV crystal.

Considering that the investigated diamond crystals have a cubic lattice belonging to the point group m3m, we define the electric field polarization direction of the laser light as an angle $\theta$ measured from the [100] crystallographic axis, with the laser beam propagating along the [001] direction. In the $xy$-plane of the crystal, the light polarization is described by $\textbf{E}=E_0\left( \hat{x}\cos(\theta) + \hat{y} \sin(\theta) \right)$. Given that all frequencies responsible for the nonlinearity are degenerate, the intrinsic permutation symmetry leaves four independent matrix elements of the $\chi^{(3)}$ tensor matrix \cite{boyd:2020, rottwitt:2015}: $\chi_{\mathrm{xxxx}}^{(3)}$, $\chi_{\mathrm{xxyy}}^{(3)}$, $\chi_{\mathrm{xyxy}}^{(3)}$ and $\chi_{\mathrm{xyyx}}^{(3)}$. For this specific geometry, the effective third-order susceptibility is written as
\begin{equation}
    \chi_{\mathrm{eff}}^{(3)}(\theta)= \chi_{\mathrm{xxxx}}^{(3)}\left ( 1- \frac{\sigma}{2} \sin^2(2\theta) \right )
    \label{eq:effective_susceptibility}
\end{equation}
where $\sigma$ is the anisotropy coefficient defined as
\begin{equation}
    \sigma=1-\frac{\chi_{\mathrm{xxyy}}^{(3)} + \chi_{\mathrm{xyxy}}^{(3)} + \chi_{\mathrm{xyyx}}^{(3)}}{\chi_{\mathrm{xxxx}}^{(3)}}.
\end{equation}
The solid curves in Fig.~\ref{fig:PolarPlot} represent the least-squares fits to Eq.(\ref{eq:effective_susceptibility}). The values of the fitted nonlinear coefficients for the studied crystals are summarized in Table II.

\begin{table}[htbp]
\centering
\small
\caption{\bf Summary of nonlinear refractive index $n_2$ and third-order susceptibility $\chi^{(3)}$  measured at 1032 nm and at $\theta=0$.}
\begin{tabular}{ccccc}
\hline
Crystal & $n_2(\theta=0)$ [$10^{-20}$ m$^2$/W]  & $ \chi_{\rm xxxx}^{(3)}$  [$10^{-22}$ m$^2$/V$^2$ ] & $ \chi_{\rm xxxx}^{(3)}$  [$10^{-14}$ esu] & $\sigma$    \\
\hline
EGSC & $3.155 (40)$ & $6.380(81)$ & $4.562(57)$ & $0.454 (57)$  \\
MCNV & $2.702 (14)$ & $5.464 (27)$ & $3.907 (20)$ & $0.418 (18)$ \\
HCNV & $2.397(39)$ & $4.847 (80)$ & $3.466 (57)$ & $0.580 (58)$ \\
\hline
\end{tabular}
  \label{tab:sample-NLC}
\end{table}
As seen in Fig.~\ref{fig:PolarPlot}, the nonlinear susceptibility or the refractive index varies with $\theta$, by approximately $50\%$, for each of the examined crystals, clearly indicating that the observed anisotropy is associated solely with diamond.\\

To explain the observed reduction in optical nonlinear susceptibility in NV-doped diamond crystals, we express the total measured susceptibility as follows: 
\begin{equation*}
    \chi_{\mathrm{eff}}^{(3)}=\chi_{\mathrm{diam}}^{(3)}+\chi_{\mathrm{NV}}^{(3)}, 
\end{equation*}
where the first term, $\chi_{\mathrm{diam}}^{(3)}$, represents the third-order susceptibility of the pristine diamond, and the second term, $\chi_{\mathrm{NV}}^{(3)}$, accounts for the contribution of NV centers.

\begin{figure}[hbt]
    \centering
    \includegraphics[width=8.0cm]{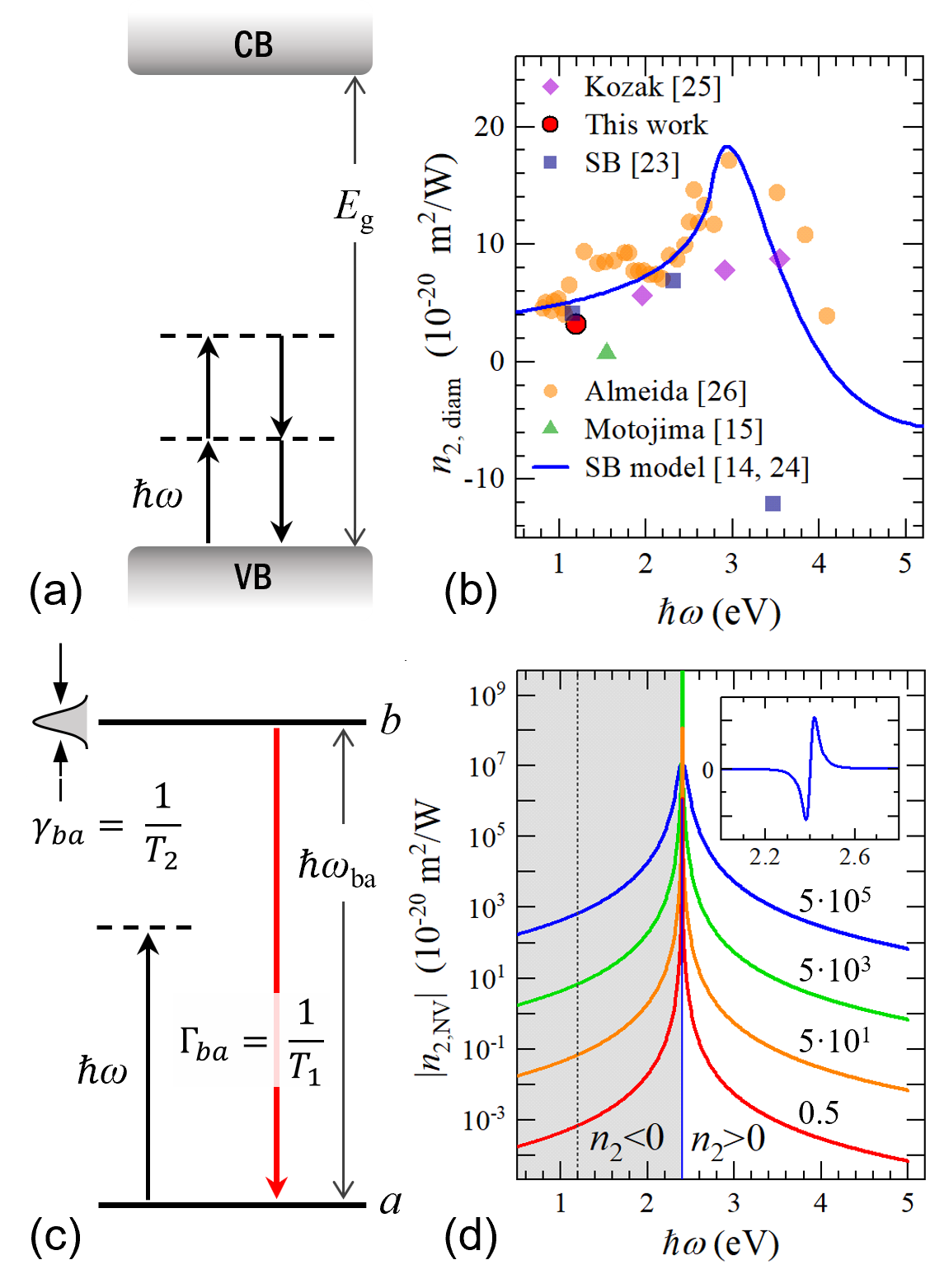}
    \caption{(a) In diamond, when the photon of energy $\hbar\omega$ is below the bandgap energy $E_{\mathrm{g}}$, the nonlinear response is primarily attributed to a parametric two-photon process. (b) Comparison of the theoretical prediction of ref. \cite{sheikbahae_1990b, sheikbahae_1991} (blue line) with various experimental results. The result of this work is represented by a red dot.
    (c) NV centers approximated as two-level systems in non-resonant laser field. $T_1$ and $T_2$ represent the effective lifetime of level $b$ and the dephasing time of the induced dipole moment. (d) The absolute value of the contribution to $n_2$ by the NV centers, calculated using the two-level atom approximation according to Eq.~(\ref{eq:n2 2-level atom}). Calculations assume $N=5.3\times 10^{22}$~m$^{-3}$, $\mu_{\mathrm{ba}}=2.23\times 10^{-29}$~$\mathrm{C\cdot m}$, and $T_1=10$~ns, with several values of $T_1/T_2$ indicated next to the corresponding curves. The resonance transition energy was assumed to be 2.4~eV ($\lambda = 517$~nm) based on the transmittance curves presented in Fig.~{\ref{Fig: crystals parameters}}(e). It should be noted that the actual values for the red-detuning (below 2.4 eV) are negative. The inset shows the linear-scale changes of $n_{2,\mathrm{NV}}$ near the resonance energy for the case $T_1/T_2=5 \cdot 10^5$.}
    \label{fig:n2_model}
\end{figure}

For the discussion of the NV effects, we apply the approach developed by Sheik-Bahae et al. for the analysis of nonlinear optical properties of semiconductors or with a wide energy band gap, like a diamond \cite{sheikbahae_1990b, sheikbahae_1991}.  They noticed that, for frequencies $\omega$ such that $\hslash\omega < E_{\mathrm{g}}$, the dominant contribution to the imaginary part of the nonlinear response arises from the instantaneous two-photon parametric absorption process, as illustrated in Fig.~\ref{fig:n2_model}(a). To determine the real part of the 3$^{\mathrm{rd}}$ order susceptibility, they applied the Kramers-Kronig transformation of the well-known imaginary (absorptive) part, which yielded the following expression \cite{sheikbahae_1990b}
\begin{equation}
    n_{2,\mathrm{diam}} \mathrm{(m^2/W}) = K'' \frac{G_2(\hslash \omega/E_{\mathrm{g}})}{n_0^2 E_{\mathrm{g}}^4},
\label{eq:n2_semiconductor}    
\end{equation}
where $K''=1.4246 \times 10^{-14}$ and $E_{\mathrm{g}}$ is given in eV, and $G_2(x)$ is the dispersion function given as: 
\begin{equation}
    G_2(x)=\frac{-2+6x-3x^2-x^3-\tfrac{3}{4}x^4-\tfrac{3}{4}x^5+2(1-2x)^{3/2}\Theta(1-2x)}{64 x^6}
\label{eq:dispersion_function}
\end{equation}
where $\Theta(x)$ represents the unit step function. Predictions of the nonlinear refractive index of solids, according to Eq.~(\ref{eq:n2_semiconductor}), are in good agreement with various experimental data as shown in Fig.~\ref{fig:n2_model}(b). Fig~\ref{fig:n2_model}(b) illustrates that $n_2$ is positive for $\hbar\omega \lesssim 0.74 E_g$, that is, for $\hbar\omega \lesssim 4.1$~eV or $\lambda \gtrsim 302.7$~nm, and negative otherwise.
Figure \ref{fig:n2_model}(b) includes data from various experiments. Although all results were obtained using the Z-scan technique, a direct comparison would be problematic due to significant variations in the experimental setups. These differences include variations in laser pulse durations, ranging from 40~fs \cite{motojima_2019} to 30~ps \cite{sheikbahae_1995}; laser wavelengths, spanning from 256~nm \cite{sheikbahae_1995} to 1500~nm \cite{almeida_2017}, as well as the properties of the investigated crystals, which were manufactured using different methods and sourced from various suppliers.\\

On the other hand, the contribution of NV color centers to $\chi_{\mathrm{eff}}^{(3)}$ can be qualitatively estimated by modeling NV$^-$ and NV$^0$ centers as a single effective two-level system interacting with a strong monochromatic laser field, as depicted in Fig.~\ref{fig:n2_model}(c). 
Despite the NV\(^-\) and NV\(^0\) centers being two different systems, their dynamics is coupled under high-intensity light pulses. This coupling results in continuous charge-state conversion NV\(^-\) $\Leftrightarrow$ NV\(^0\), through multiphoton ionization and light-induced recombination \cite{motojima_2019, ji_2018}. Importantly, despite these charge-state conversions, the total population of NV centers remains conserved. Furthermore, based on the broad transmission spectra shown in Figs.~\ref{Fig: crystals parameters}(e, f), we interpret the states \(a\) and \(b\) as effective states representing the ground ($^3A_2$ and $^2E$) and excited ($^3E_2$ and $^2A$) states for both types of both NV$^-$ and NV$^0$ centers, respectively, and encompassing their associated phonon structures.

Following Boyd's formulation \cite{boyd:2020}, the optical susceptibility of such two-level system, parametrized by the saturation field strength $|E_s^0|^2=\hbar^2/(4 |\mu_{\mathrm{ba}}|^2 T_1 T_2)$, is described by
\begin{equation}
    \chi=-N |\mu_{\mathrm{ba}}|^2 \frac{T_2}{\epsilon_0 \hbar} \frac{\Delta T_2 - i}{1+\Delta^2 T_2^2 +|E|^2/|E_s^0|^2},
\end{equation}
where $\Delta=\omega-\omega_{\mathrm{ba}}$ represents the detuning (with $\omega_{\mathrm{ba}}$ being the resonance frequency), $\mu_{\mathrm{ba}}$ is the induced dipole moment, and $N$ stands for the number density of NV centers. $T_1$ and $T_2$ correspond to the  effective lifetime of level $b$ and the characteristic dephasing time of the dipole moment, respectively. 

By performing a power series expansion in $|E|^2/|E_s^0|^2$ and comparing the result with the standard power series expansion $\chi=\chi^{(1)}+3 \chi^{(3)} E E +\dots\;$ where $\chi^{(3)}(\omega)\equiv \chi^{(3)}(\omega=\omega+\omega-\omega)$, that is, in the degenerate case we obtain the third-order nonlinear susceptibility. Using the above expansion together with Eq.~(\ref{eq:n2 to chi^3}), we arrive at the following expression for the nonlinear refractive index: 
\begin{equation}
    n_{2,\mathrm{NV}}= N |\mu_{\mathrm{ba}}|^4 
    \frac{T_1 T_2^2}{n_0^2 \epsilon_0^2 c \hslash^3} \frac{\Delta T_2}{\left ( 1+\Delta^2 T_2^2 \right )^2}.
    \label{eq:n2 2-level atom}
\end{equation}
 In the non-resonant limit $|\Delta T_2| \gg 1$, this simplifies to the form:
 \begin{equation}
     n_{2,\mathrm{NV}} \cong N \frac{|\mu_{\mathrm{ba}}|^4}{n_0^2 \epsilon_0^2 c \hslash^3} \frac{1}{\Delta^3} \frac{T_1}{T_2}.
\end{equation}
 As follows from the above equations, in the case of red-detuning ($\Delta<0$), the contribution of the two-level system to the nonlinear refractive index is negative, while for blue-detuning ($\Delta>0$), it is positive. 
 In our case, based on the experimental data presented in Fig.~\ref{Fig: crystals parameters}(e,f), we assumed that we are dealing with a two-level system with a resonance transition wavelength of approximately 520~nm. Consequently, the wavelength of our laser is red-detuned relative to this transition, and therefore NVs give a negative contribution to $n_2$. 
 
 Although the qualitative picture is clear, determining the absolute values of $n_{2,\mathrm{NV}}$ is problematic due to the lack of data on $T_1$ and $T_2$, as well as the dipole moment $\mu_{\mathrm{ba}}$ of the considered transition. 
 Figure \ref{fig:n2_model}(d) shows the absolute values of $n_{2,\mathrm{NV}}$ as a function of the energy detuning of the laser pulse relative to the energy of the resonance transition considered (2.4 eV), calculated for different values of $T_1/T_2$. In the calculations, we assumed $N=5.3\times 10^{22}$~$\mathrm{m^{-3}}$ which represents the concentration of NV in the MCNV crystal, while $T_1$ is approximately 10~ns which, according to the Einstein relation, corresponds to $\mu_{\mathrm{ba}}=2.23\times 10^{-29}$~$\mathrm{C\cdot m}$. The curves in the figure were calculated for an increasing ratio $T_1/T_2$, ranging from 0.5 to $5\times 10^{5}$. The lowest ratio corresponds to a radiatively broadened transition, while the highest is estimated on the basis of the bandwidth of the absorbance curve of several dozen nanometers, corresponding to $T_2$ of a few dozen femtoseconds. For detuning of approximately 1.2~eV, $|n_{\mathrm{2, NV}}|$ can be as high as about $6 \times 10^{-18}$~$\mathrm{m^2\cdot W^{-1}}$, demonstrating that NV color centers modeled as two-level systems, can actually explain the observed reduction in the nonlinear refractive index of diamonds doped with NV within the studied spectral range. Moreover, it is very likely that this quantity varies significantly with detuning, which, however, requires more realistic modeling of phonon-broadened optical transitions as well as further experimental verification employing tunable femtosecond laser sources.

\section{Summary and Conclusions}
In this work, we investigated the nonlinear refractive index ($n_2$) of NV-doped diamond crystals via the optical Kerr effect. The OKE was studied using the Z-scan technique with 230-fs laser pulses at 1032 nm -- twice the wavelength commonly used for NV center excitation.

Our results reveal that, for all crystals examined (both undoped and NV-doped), $n_2$ is positive and exhibits anisotropy characterized by a four-fold rotational symmetry. To our knowledge, this is the first observation of such behavior. 
Furthermore, the magnitude of $n_2$ decreases with NV concentration and is lower in NV-doped diamonds compared to pure diamond. This reduction is attributed to the negative contribution of NV centers to the third-order susceptibility, modeled as a two-level system perturbed by a strong laser field red-detuned from their effective resonance frequency.

These findings represent only the beginning of optical nonlinear studies on volume-doped NV diamonds. Future work will focus on quantitative modeling of NV-laser interactions, particularly under ultra-short laser pulses, as well as systematic exploration of how nonlinear optical properties depend on laser wavelength and pulse duration. Nonlinear absorption effects, such as saturable and multiphoton absorption, appear significant and are currently under our investigation. Additionally, examining the roles of other defects and crystal orientations could yield insights into the tailoring of optical properties for specific applications. \\

The optical Kerr effect in NV-doped diamonds holds promise for ultrafast optical switching and modulation. Moreover, multiphoton excitation could enable advances in quantum coherence studies, energy-transfer mechanisms, and high-resolution optical microscopy. Investigating nonlinear optical effects in volume-doped NV diamonds is a vital step toward unlocking their potential for next-generation photonic and quantum technologies.\vspace{0.75cm}

\begin{backmatter}

\bmsection{Funding}
This research was funded in part by National Science Centre, Poland grant 2021/03/Y/ST3/00185.\\ For the purpose of Open Access, the author has applied a CC-BY public copyright licence to any Author Accepted Manuscript (A  AM) version arising from this submission.

\bmsection{Acknowledgments}
The study was carried out using research infrastructure purchased with the funds of the European Union in the framework of the Smart Growth Operational Programme, Measure 4.2; Grant No. POIR.04.02.00-00-D001/20, “ATOMIN 2.0 - ATOMic scale science for the INnovative economy”.

\bmsection{Disclosures}
The authors declare no conflicts of interest.

\bmsection{Data Availability Statement}
Data underlying the results presented in this paper are available in Ref. [x] (after review)

\end{backmatter}

\bibliography{bib}

\end{document}